\documentclass[12pt]{article}

\newcommand{\be}{\begin{eqnarray}}
\newcommand{\ee}{\end{eqnarray}}

\begin{document}

\rightline{\Large{Preprint RM3-TH/04-13}}

\vspace{1cm}

\begin{center}

\LARGE{Generalized Parton Distributions\\ in the Light-Front Constituent Quark Model\footnote{\bf To appear in the Proceedings of the International Conference on {\em The Physics of excited Nucleons} ($N^*~2004$), Grenoble (France), March 24-27, 2004, World Scientific Publishing (Singapore), in press.}}\\

\vspace{1cm}

\large{Silvano Simula}\\

\vspace{0.5cm}

\normalsize{Istituto Nazionale di Fisica Nucleare, Sezione di Roma III,\\ Via della Vasca Navale 84, I-00146 Roma, Italy\\ E-mail: simula@roma3.infn.it}

\end{center}

\vspace{1cm}

\begin{abstract}

\noindent The Generalized Parton Distributions ($GPD$s) of the nucleon are analyzed within the relativistic constituent quark model formulated on the light-front. It is shown that the matrix elements of the {\em plus} component of the one-body vector current are plagued by {\em spurious} effects related to the dependence on the hyperplane where the nucleon wave function is defined in terms of its constituents. The {\em physical} $GPD$s can be extracted only from the matrix elements of a {\em transverse} component of the one-body current. The loss of the polinomiality property is then related to the neglect of the pair creation process for non-vanishing values of the skewness. The need of implementing effective many-body currents corresponding to the $Z$-graph is stressed and a possible approach to achieve such a goal is proposed.

\end{abstract}

\newpage

\pagestyle{plain}

\section{Introduction}

The physics issue of Generalized Parton Distributions ($GPD$s) is attracting a lot of theoretical interest and its experimental investigation is now included in the plans of various laboratories in the world. The importance of $GPD$s relies on the fact that they represent a unified theoretical picture of various inclusive and exclusive (both polarized and unpolarized) processes off hadrons in the Deep Inelastic Scattering ($DIS$) regime. The $GPD$s provide information of the longitudinal and transverse distributions of partons inside the hadrons, because they are off-forward matrix elements of quark and gluon operators~\cite{Diehl}. The $GPD$s depend on three variables: the internal light-front ($LF$) fraction $x$, the squared 4-momentum transfer $\Delta^2$ and the skewness $\xi$, which is the relative change of the $LF$ fraction of the hadron momentum. In the forward limit ($\xi = \Delta^2 = 0$) the $GPD$s reduce to the usual Parton Distribution Functions ($PDF$s), while their integrals over $x$ provide the elastic form factors of the hadron under investigation.

Using deeply virtual Compton scattering processes as well as hard exclusive production of vector mesons, information on the $GPD$s may be extracted from data (see~\cite{data}). Therefore various calculations of $GPD$s have been already performed using different models of the hadronic structure, like the bag model \cite{bag}, the chiral quark-soliton model \cite{soliton} and the Constituent Quark ($CQ$) model \cite{CQM,LFCQM}. We should mention also phenomenological estimates of the $GPD$s, obtained using parameterizations of the usual $PDF$s and factorizing the $\Delta^2$-dependence according to the elastic form factors \cite{phen}.

In this contribution we consider the estimate of the nucleon $GPD$s within the relativistic $CQ$ model formulated on the light-front. Our aim is to extend to the case of $GPD$s some of the results of the analysis made in Refs.~\cite{FS,SIM02} on elastic hadron form factors. It will be shown that the matrix elements of the {\em plus} component of the one-body vector current are plagued by {\em spurious} effects related to the dependence on the hyperplane where the nucleon wave function is defined in terms of its constituents. The {\em physical} $GPD$s can be extracted only from the matrix elements of a {\em transverse} component of the one-body current. The loss of the polinomiality property is then related to the neglect of the pair creation process ($Z$-graph \cite{Zgraph}) for non-vanishing values of $\xi$. The need of implementing effective many-body currents corresponding to the $Z$-graph will be stressed, and a possible approach to achieve such a goal will be proposed.

\section{Definition of nucleon $GPD$s and link with the $CQ$ model}

In what follows we limit ourselves to the case of the nucleon $GPD$s related to the vector current, adopting the so-called symmetric frame. According to Ref.~\cite{Ji} the nucleon off-forward, twist-two parton distributions are defined in terms of light-cone correlation functions in the gauge $A \cdot \omega = 0$ as
 \be
    \int {d\lambda \over 2 \pi} e^{i \lambda x} \langle P^{\prime} \nu^{\prime} 
    | \bar{\psi}(- \lambda \omega / 2) \gamma^{\mu} \psi(\lambda \omega / 2) 
    | P \nu \rangle = u(P^{\prime} \nu^{\prime}) ~ \Gamma^{\mu} ~ u(P \nu) + 
    ... ~ ,
    \label{eq:def}
 \ee
where $\omega$ is a null vector, $\nu$ ($\nu^{\prime}$) is the initial (final) helicity of the nucleon, the dots denote higher-twist terms and
 \be
    \Gamma^{\mu} = H(x, \xi, \Delta^2) ~ \gamma^{\mu} + E(x, \xi, \Delta^2) ~ 
    {i \sigma^{\mu \rho} \Delta_{\rho} \over 2M} 
    \label{eq:HE}
 \ee
with $\Delta \equiv P^{\prime} - P$ being a space-like 4-vector ($\Delta^2 \leq 0$), $\xi \equiv -\omega \cdot \Delta / (2 \omega \cdot \overline{P})$ and $\overline{P} = (P^{\prime} + P)/ 2$. The physical support of the $GPD$s is $|x| \leq 1$ (see~\cite{Diehl}).

The nucleon $GPD$s, $H(x, \xi, \Delta^2)$ and $E(x, \xi, \Delta^2)$, can be extracted from the matrix elements of the {\em plus} component of the vector current. Denoting $[ u(P^{\prime} \nu^{\prime}) ~ \Gamma^{\mu} ~ u(P \nu) ~ / ~ 2 \overline{P}^+ ]$ by $I_{\nu^{\prime} \nu}^{\mu}$, one has
 \be
    I_{\nu^{\prime} \nu}^+ = \sqrt{1 - \xi^2} \left[ H - {\xi^2 \over 1 - 
    \xi^2} E \right] \delta_{\nu^{\prime} \nu} - i {\sqrt{\Delta_0^2 - 
    \Delta^2} \over 2M} ~ E ~ (\sigma_2)_{\nu^{\prime} \nu} 
    \label{eq:plus}
 \ee
where  $\sigma_2$ is a Pauli matrix and $\Delta_0^2 \equiv - 4M^2 \xi^2 / (1 - \xi^2)$. In Eq.~(\ref{eq:plus}) the transverse vector $\vec{\Delta}_{\perp}$ is assumed to be $(\Delta_{\perp}, 0)$ with $\Delta_{\perp} = \sqrt{(1 - \xi^2) (\Delta_0^2 - \Delta^2)}$.

In the forward limit $\Delta^{\rho} \to 0$ one has $H(x, 0, 0) = q(x_B = x)$ for $x > 0$ and $H(x, 0, 0) = -\bar{q}(x_B = -x)$ for $x < 0$, where $x_B$ is the Bjorken variable and $q(x_B)$ is the ordinary $PDF$, while 
 \be
    \int_{-1}^{1} dx ~ H(x, \xi, \Delta^2) = F_1(\Delta^2) , \qquad
    \int_{-1}^{1} dx ~ E(x, \xi, \Delta^2) = F_2(\Delta^2) ,
    \label{eq:SR}
 \ee
where $F_1$ ($F_2$) is the Dirac (Pauli) elastic form factor of the nucleon. Note that the l.h.s.~of Eq.~(\ref{eq:SR}) is independent on the skewness $\xi$.

An estimate of the nucleon $GPD$s has been recently proposed in Ref.~\cite{LFCQM} using the relativistic $CQ$ model formulated on the light-front, at least for large values of $x$ where valence quark degrees of freedom dominate the behavior of $DIS$ structure functions. In~\cite{LFCQM} the expansion of the nucleon state in terms of $N$-parton $LF$ wave functions, derived in Ref.~\cite{LFWFs}, was used to argue a correspondence (or link) between the nucleon wave function pertinent to relativistic $CQ$ models and the lowest Fock-space component of the nucleon, i.e.~the valence-quark $LF$ wave function. As a result the valence-quark contribution to the $GPD$s was explicitly obtained~\cite{LFCQM} in terms of the three-$CQ$ wave function of the nucleon arising in quark potential models. The calculated $GPD$s have however a restricted physical support, $x \geq \xi$, and, moreover, their first moments (\ref{eq:SR}) do depend upon $\xi$.

\section{Spurious effects in the $LF$ $CQ$ model}

Two are the main criticisms that we make to the link proposed in~\cite{LFCQM}. The first one is that the $CQ$s are not expected at all to be in a one-to-one correspondence with current quarks, like the valence quarks, at least because of the different mass. Moreover, the hypothesis of a compositeness of the $CQ$s has a very long history, and we limit ourselves just to mention the recent analysis made in~\cite{Petronzio}, where an almost model-independent evidence for extended substructures in the proton was found in the low-$\Delta^2$ data on the structure function $F_2$ of the proton. As pointed out in~\cite{CQM}, the assumption that $CQ$s have their own partonic content, allows to generate $GPD$s with a physical support not restricted to $x \geq \xi$, but corresponding at least to $x \geq - \xi$. This is not a trivial point, but in order to better focus on the second criticism the $CQ$ compositeness will be neglected hereafter.

The second criticism arises from the fact that, following the link proposed in~\cite{LFCQM}, one is led to estimate the valence-quark contribution to the $GPD$s using the one-body approximation for the vector current operator. While in field theory current operators are one-body, in relativistic quantum models, like the $LF$ one adopted in~\cite{LFCQM}, the restriction to a finite number of degrees of freedom and the requirement of Poincar\`e covariance introduce unavoidably many-body terms in the current operators (see~\cite{LF}). Therefore, the $CQ$ estimates of the $GPD$s should be extracted from the matrix elements of a complicated many-body operator, which we denote by 
 \be
    \overline{I}_{\nu^{\prime} \nu}^{\mu}(x, \xi, \Delta^2) \equiv \langle 
    \Psi^{(LF)}(P^{\prime} \nu^{\prime}) | ~ \overline{V}^{\mu} ~ | 
    \Psi^{(LF)}(P \nu) \rangle ~ / ~ 2 \overline{P}^+
    \label{eq:LF}
 \ee
where $\Psi^{(LF)}$ denoted the $LF$ (three-$CQ$) wave function of the nucleon as described in~\cite{FS,LFCQM}, and $\overline{V}^{\mu} = \sum_j e_j ~  \gamma^{\mu} ~ \delta[x + \xi - (1 + \xi) ~ x_j] + (many-body ~ terms)$, with $x_j$ being the $LF$ fraction of the struck $CQ$. The matrix elements (\ref{eq:LF}) have the same Lorentz decomposition as in Eq.~(\ref{eq:HE}), viz.
 \be
    \overline{I}_{\nu^{\prime} \nu}^{\mu} & = & u(P^{\prime} \nu^{\prime}) ~ 
    \overline{\Gamma}^{\mu} ~ u(P \nu) ~ / ~ 2 \overline{P}^+ ~ , \nonumber \\
    \overline{\Gamma}^{\mu} & = & \overline{H}(x, \xi, \Delta^2) ~ \gamma^{\mu} 
    + \overline{E}(x, \xi, \Delta^2) ~ {i \sigma^{\mu \rho} \Delta_{\rho} 
    \over 2M} ~ ,
    \label{eq:LF_full}
 \ee
where $\overline{H}$ and $\overline{E}$ are the $CQ$ estimates of the nucleon $GPD$s. Thus the latter can be extracted from the matrix elements of the {\em plus} component, $\overline{I}_{\nu^{\prime} \nu}^+$, as in Eq.~(\ref{eq:plus}). Note that if the Lorentz decomposition (\ref{eq:LF_full}) holds, then the same $GPD$s can be extracted also from the transverse matrix elements with $\mu = y$, where $y$ is the direction transverse to $\vec{\Delta}_{\perp}$ in the $\perp$-plane.

The critical point occurs when the full current $\overline{V}^{\mu}$ is replaced by its one-body approximation
 \be
    \overline{V}^{\mu} \to \overline{V}_{(1)}^{\mu} = \sum_j e_j ~ 
    \gamma^{\mu} ~ \delta[x + \xi - (1 + \xi) ~ x_j] ~ .
    \label{eq:one-body}
 \ee
As firstly pointed out by Karmanov and Smirnov~\cite{Karmanov} in case of form factors and subsequently derived from the analysis of the Feymann triangle diagram in Ref.~\cite{DS}, the Lorentz decomposition of the matrix elements of an approximate current includes {\em spurious} structures depending on the null 4-vector $\omega^{\mu}$, which identifies the orientation of the hyperplane where $LF$ wave functions are defined. The particular direction defined by $\omega^{\mu}$ is irrelevant in the possible Lorentz structures only when the full current $\overline{V}^{\mu}$ is used. In other words, in full analogy with the case of the elastic nucleon form factors analyzed in Refs.~\cite{Karmanov_N,FS}, one has
 \be
    \overline{I}_{\nu^{\prime} \nu}^{\mu} & \to & \left( \overline{I}_{(1)} 
    \right)_{\nu^{\prime} \nu}^{\mu} = u(P^{\prime} \nu^{\prime}) ~ \left\{ 
    \overline{\Gamma}_{(1)}^{\mu} + \overline{\mathcal{B}}_{(1)}^{\mu}\right\} 
    ~ u(P \nu) ~ / ~ 2 \overline{P}^+ ~ , \nonumber \\
    \overline{\Gamma}_{(1)}^{\mu} & = & \overline{H}_{(1)}(x, \xi, \Delta^2) ~ 
    \gamma^{\mu} + \overline{E}_{(1)}(x, \xi, \Delta^2) ~ {i \sigma^{\mu \rho} 
    \Delta_{\rho} \over 2M} ~ , \nonumber \\
    \overline{\mathcal{B}}_{(1)}^{\mu} & = & \overline{B}_{(1)}(x, \xi, 
    \Delta^2) \left[ {\gamma \cdot \omega \over \omega \cdot \overline{P}} - 
    {1 \over M (1 + \eta)} \right] \overline{P}^{\mu} \nonumber \\
    & + & \overline{C}_{(1)}(x, \xi, \Delta^2) {\omega^{\mu} \over \omega 
    \cdot \overline{P}} + \overline{D}_{(1)}(x, \xi, \Delta^2) {\gamma \cdot 
    \omega \over \omega \cdot \overline{P}} {\omega^{\mu} \over \omega \cdot 
    \overline{P}} ~ ,
    \label{eq:LF_one-body}
 \ee
where $\overline{B}_{(1)}$, $\overline{C}_{(1)}$ and $\overline{D}_{(1)}$ are {\em spurious} $GPD$s, while $\overline{H}_{(1)}$ and $\overline{E}_{(1)}$ are the {\em physical} $GPD$s in the one-body approximation; moreover, $\eta \equiv -\Delta^2 / 4 M^2$. Note that $\Delta_{\mu} \overline{\mathcal{B}}_{(1)}^{\mu} \propto \xi$, so that gauge-invariance is lost for $\xi \neq 0$.

The presence of the $\omega$-dependent structures in Eq.~(\ref{eq:LF_one-body}) does not allow any more to extract the physical $GPD$s from the matrix elements of the {\em plus} component of the vector current. Indeed one has
 \be
    \left( \overline{I}_{(1)} \right)_{\nu^{\prime} \nu}^+ & = & \sqrt{1 - 
    \xi^2} \left[ \widetilde{H}_{(1)} - {\xi^2 \over 1 - \xi^2} 
    \widetilde{E}_{(1)} \right] \delta_{\nu^{\prime} \nu} - 
    i {\sqrt{\Delta_0^2 - \Delta^2} \over 2M} \widetilde{E}_{(1)} 
    (\sigma_2)_{\nu^{\prime} \nu} ~ , \nonumber \\
    \widetilde{H}_{(1)} & = & \overline{H}_{(1)} -{\xi^2 \over 1 - \xi^2} 
    \overline{E}_{(1)} + \overline{B}_{(1)} \left[ 1 - {1 \over (1 + \eta) 
    (1 - \xi^2)} \right] ~ , \nonumber \\
    \widetilde{E}_{(1)} & = & \overline{E}_{(1)} + \overline{B}_{(1)} / 
    (1 + \eta) ~ .
    \label{eq:HE_plus}
 \ee
Note that $\widetilde{H}_{(1)}(x, 0, 0) = \overline{H}_{(1)}(x, 0, 0)$, i.e.~forward parton distributions are free from spurious effects, as it should be in the Bjorken regime.

The matrix elements of the $y$ component of the vector current are free from spurious effects and the {physical} $GPD$s can be obtained from
 \be
    \left( \overline{I}_{(1)} \right)_{\nu^{\prime} \nu}^y & = & i {M \over 
    \overline{P}^+} {1 \over \sqrt{1 - \xi^2}} \left\{ - \xi \left[ 
    \overline{H}_{(1)} - {\Delta_{\perp}^2 \over 4 M^2} \overline{E}_{(1)} 
    \right] (\sigma_1)_{\nu^{\prime} \nu} \right. \nonumber \\
    & + & \left. {\Delta_{\perp} \over 2M} \left[ \overline{H}_{(1)} + 
    \overline{E}_{(1)} \right] (\sigma_3)_{\nu^{\prime} \nu} \right\}
    \label{eq:y}
 \ee
with $\Delta_{\perp} = \sqrt{(1 - \xi^2) (\Delta_0^2 - \Delta^2)}$. Similar results holds as well in case of the helicity dependent $GPD$s related to the axial current. We do not report here the final formulae for lack of space.

Any difference between $\widetilde{H}_{(1)}$ ($\widetilde{E}_{(1)}$) and $\overline{H}_{(1)}$ ($\overline{E}_{(1)}$) is the signature of spurious effects, which can therefore affect the dependence of the model $GPD$s on $x$, $\xi$ and $\Delta^2$. The numerical impact in case of the explicit calculations of Ref.~\cite{LFCQM} is left to future works. We just mention that in case of the magnetic form factors of the nucleon the very existence and relevance of spurious effects have been already demonstrated in Ref.~\cite{FS} (see there Fig.~5). Another interesting case, in which the role of spurious effects shows up dramatically, is represented by the nucleon axial charge, $g_A$. In Ref.~\cite{Brodsky} it is claimed that the difference between the experimental value ($g_A^{(exp.)} = 1.26$) and the non-relativistic $CQ$ result ($g_A^{(NR)} = 1.67$) can be explained by relativistic effects due to Melosh rotations. However, the $LF$ calculation of Ref.~\cite{Brodsky} is plagued by spurious effects and therefore it is not correct. Once the spurious effects are properly subtracted and the same values of the relevant $CQ$ parameters used in~\cite{Brodsky} are adopted, the $LF$ estimate becomes $g_A = 1.63$, i.e.~quite close to the non-relativistic result~!

\section{Pair creation process from the vacuum}

The use of the $y$ component of the vector current allows to obtain nucleon $GPD$s free from spurious effects. However, an important drawback of the one-body approximation (\ref{eq:one-body}) still remains, namely the loss of the polinomiality property~\cite{Ji} or, simply, the fact that the first moments of the model $GPD$s do depend upon $\xi$. Such a dependence is nothing else than the frame-dependence of the form factors calculated within the one-body current. In this respect an extensive investigation of the elastic form factors of pseudoscalar and vector two-fermion systems has been carried out in Ref.~\cite{SIM02}, adopting the $LF$ Hamiltonian formalism both at $\xi = 0$ and $\xi \neq 0$. Huge numerical differences in the form factors have been found between the frame in which $\xi = 0$ and the one where $\xi = \xi_{max} = \sqrt{\Delta^2 / (\Delta^2 - 4 M^2)}$. The origin of such differences is entirely due to the pair creation process from the vacuum ($Z$-graph), as shown in Ref.~\cite{SIM02} using the results of the analysis of the Feynman triangle diagram obtained in Ref.~\cite{DS}.

The $Z$-graph is a process beyond the one-body approximation and represents a many-body current, which vanishes at $\xi = 0$ and becomes more and more important as $\xi$ increases. Its evaluation is mandatory to obtain the appropriate $\xi$-dependence of the $GPD$s, so that the form factors obtained by integration of the $GPD$s become truly frame independent.

Since the contribution of the $Z$-graph is vanishing when $\xi = 0$, the form factors obtained using the one-body current plus the $Z$-graph at $\xi \neq 0$ should coincide with the ones calculated directly at $\xi = 0$ with the one-body term only. Thus, within the $LF$ formalism the evaluation of the $Z$-graph is not strictly necessary to calculate the form factors. These can be obtained by choosing $\xi = 0$, which kills the $Z$-graph and minimizes therefore the impact of many-body currents (see~\cite{SIM02}). The vanishing of the $Z$-graph at $\xi = 0$ is peculiar of the $LF$ formalism. Indeed, in other Dirac forms of the dynamics, like the point-form, it is not possible to find a frame where the $Z$-graph is vanishing. Thus the impact of many-body currents may become quite important in the point form, as it has been extensively investigated by Desplanques (see Ref.~\cite{Desplanques} and references therein).

The explicit construction of an effective many-body current corresponding to the $Z$-graph is important for two reasons: i) to obtain form factors which are independent (at least to a large extent) of the particular choice of the form of the dynamics; ii) to estimate the appropriate $\xi$-dependence of $GPD$s fulfilling the polinomiality property. Such a construction is not an easy task. Up to our knowledge, the most promising approach to include the effects of the $Z$-graph is the dispersion formulation of the $CQ$ model~\cite{Dima}. Such a formulation allows a covariant evaluation of the Feynman triangle diagram (including the $Z$-graph), providing therefore hadron form factors which are exactly frame independent. As expected, the dispersion result coincides with the $LF$ one at $\xi = 0$. The dispersion $CQ$ model has been also applied~\cite{MNS} to time-like 4-momentum transfers ($\Delta^2 > 0$), where the usual $LF$ formalism (see~\cite{Bmeson}) is plagued by the neglect of the $Z$-graph.

The dispersion formulation of the $CQ$ model does already the requested job of including the effects of the $Z$-graph in case of form factors. Its extension to off-forward matrix elements and the inclusion of the $CQ$ compositeness represent in our opinion the most promising way to achieve a covariant $CQ$ calculation of hadron $GPD$s.

\section{Conclusions}

The nucleon generalized parton distributions have been analyzed within the relativistic constituent quark model formulated on the light-front. We have shown that the matrix elements of the {\em plus} component of the one-body vector current are plagued by {\em spurious} effects related to the dependence on the hyperplane where the nucleon wave function is defined in terms of its constituents. The {\em physical} $GPD$s can be extracted only from the matrix elements of a {\em transverse} component of the one-body current. The loss of the polinomiality property is related to the neglect of the pair creation process for non-vanishing values of the skewness. The need of implementing effective many-body currents corresponding to the $Z$-graph is stressed, and in this respect the use of the dispersion formulation of the constituent quark model~\cite{Dima} is expected to be the most promising way to achieve covariant $CQ$ estimates of the nucleon $GPD$s fulfilling the polinomiality property.

\end{document}